**Exploring the relationship between technological improvement and innovation diffusion: An empirical test**

JongRoul Woo[a,*] and Christopher L. Magee[a,b]

[a] *Institute for Data, Systems, and Society, Massachusetts Institute of Technology, 77 Massachusetts Avenue, Cambridge, MA 02139-4307, United States*

[b] *SUTD-MIT International Design Center, Massachusetts Institute of Technology, 77 Massachusetts Avenue, Cambridge, MA 02139-4307, United States*

[*] Corresponding author. E-mail: jroul86@mit.edu; Tel.: +1-617-386-3392

**Abstract**

Different technological domains have significantly different rates of performance improvement. Prior theory indicates that such differing rates should influence the relative speed of diffusion of the products embodying the different technologies since improvement in performance during the diffusion process increases the desirability of the product diffusing. However, there has not been a broad empirical attempt to examine this effect and to clarify the underlying cause. Therefore, this paper reviews the theoretical basis and focuses upon empirical tests of this effect across multiple products and their underlying technologies. The results for 18 different diffusing products show the expected relationship-faster diffusion for products based on more rapidly improving technological domains- between technological improvement and diffusion with strong statistical significance. The empirical examination also demonstrates that technological improvement does not slow down in the latter parts of diffusion when penetration does slow down. This finding indicates that diffusion slow down in the latter stages is due to market saturation effects and is not due to slowdown of performance improvement.

# 1. Introduction

It is widely accepted that technological change is a major source of economic growth (Romer, 1990; Solow, 1957). Although invention and innovation are essential aspects of technological change, diffusion is critical to economic or social impact since innovations must spread across their potential markets over time to have such impact. Because of this importance of the diffusion process in technological change, there have been many technological change studies geared toward understanding diffusion.

First, it has been observed that the diffusion pattern of successful innovations over time generally follows an S-curve, but diffusion speeds (In general, the speed indicates the distance traveled divided by the travel time. In the diffusion context, the distance can be defined as the difference between two penetration levels and the time can be defined as the amount of time it takes to go from one penetration level to another) at similar phases of the process vary considerably for different innovations (Geroski, 2000; Hall, 2005). Other studies have identified the factors affecting diffusion speeds and explaining variation in diffusion speeds for different innovations. Some of these studies have taken a static view of diffusing innovations and emphasized the effects of economic and social environment factors on diffusion speeds (Bayus, 1992; Cho et al., 2012; Griliches, 1957; Lee et al., 2017; Mansfield, 1961; 1989; Olshavsky, 1980; Rogers, 1995; Van den Bulte, 2000). Meanwhile, there are theoretical studies noting that technological improvement also significantly affects the diffusion process (Chow, 1967; Davies, 1979; Ireland and Stoneman, 1986; Metcalfe, 1981; Rosenberg, 1976; Stoneman and Ireland, 1983; Stoneman, 2002) and empirical studies to provide evidence for this effect in some products such as computers, numerically controlled machine tools, semi-dwarf wheat, and telephones (Bagchi et al, 2008; Chow, 1967; Karshenas and Stoneman, 1993; Knudson, 1991; Stoneman and Toivanen, 1997). However, none of the empirical studies look at multiple domains with different rates of improvement- our focus.

In this regard, it is well established that the performance of all technological domains that have been measured increase exponentially over time by different rates (Farmer and Lafond, 2016; Koh and Magee, 2006; 2008; Magee et al., 2016; Moore, 1965; Nagy et al., 2013; Sahal, 1979). Based on the previous studies mentioned above, one might intuitively speculate that such differing rates will influence the relative speed of diffusion and an initial diffusion speed would be more strongly accelerated for an innovation that improves more rapidly. However, there has not been a cross-domain empirical study. Having such a quantitative basis will allow one to estimate the benefits for diffusion of more rapidly improving performance. This paper relates theoretical and empirical studies by reviewing the theoretical basis and focusing upon empirical tests of this effect across multiple products and their underlying technologies.

As a result, this study empirically shows that new products that are based on faster (slower) improving technological domains are spread more rapidly (slowly) in the market and the intensity of this relationship becomes weaker toward the later stage of the diffusion process. It also finds that technological improvement does not slow down in the latter parts of diffusion when penetration does slow down. Our findings add to the limited empirical evidence on the relationship of technological improvement with diffusion and clarify the underlying cause of this relationship.

The remainder of this paper is organized as follows. Section 2 describes our theoretical framework and develops the hypotheses to be tested. Section 3 presents the data and methods used in this study. Section 4 presents empirical results on the relationship of technological improvement and innovation diffusion. Section 5 interprets the results and discusses their implications. Section 6 provides conclusions.

## 2. Theory and hypotheses

Products are designed to perform specific functions based on related technological domains and are then released into the market. For example, the automobile is a transportation artifact based on the internal combustion engine as a key technology, and the mobile phone is a communication device based on wireless telecommunication technology as a key technology. In each of the underlying technological domains, there are inventions that drive the technological improvement (Benson and Magee, 2015) and we investigate the relationship between diffusion speed of the products and improvement rate of their key technological domains.

It is well known that the adoption of new products over time follows an S-curve and diffusion speeds vary across products (Geroski, 2000; Hall, 2005). There are two theoretical frameworks for modeling the S-curve of diffusion. First, the most widely used model is based on epidemic theories, which assume that consumers have the same taste and that the new product is constant over time (Geroski, 2000; Griliches, 1957; Mansfield, 1961). In this epidemic diffusion model, people adopt the new product through the influence of existing adopters and more and more consumers adopt the product as time passes. This model predicts that diffusion of a product follows a logistic function with three parameters, the inflection point, the asymptote, and the steepness of the S-curve (Figure 1A shows these parameters on a schematic S-curve). The parameters are related with the timing, the maximum number of potential adopters, and the overall speed of the diffusion process. In addition, there is an extended epidemic diffusion model called the stock adjustment model, which assumes that the product is not constant over time and technological improvement affects the maximum number of potential adopters by widening the potential market for new product (Chow, 1967; Karshenas and Stoneman, 1993). Moreover, an alternative model assumes that consumers' tastes follow a normal distribution and the performance and price of new product changes over time (Davies, 1979; Geroski, 2000). This model assumes that consumers adopt the new product when their utility for this product exceeds their threshold level. Each of these theoretical frameworks reinforces the S-curve of the diffusion process and essentially accounts for the decreasing diffusion speed (or second part of the S) by market saturation.

Such theoretical frameworks suggest that accounting for technological improvement during the diffusion process would have an effect. Indications of this effect can be gleaned through some empirical studies that investigated the effect of increased performance (quality adjusted price) on the diffusion process of computers (Chow, 1967), numerically controlled machine tools (Stoneman and Toivanen, 1997), semi-dwarf wheat (Knudson, 1991), and telephones (Bagchi et al., 2008). However, the empirical evidence is limited in each case to a single product and the systemic effect of improving technological performance on the diffusion process across a set of products and across their key technological domains has not been studied empirically. In addition to the theoretical arguments that technological improvement promotes the adoption of new product, there are also studies that argue that consumers' expectation of technological improvement may delay the adoption of new product (Ireland and Stoneman, 1986; Karshenas and Stoneman, 1993; Rosenberg, 1976). However, in this paper, we do not discuss this argument in detail because this effect is difficult to be confirmed in the design of this study.

There are empirical studies finding factors to explain variation in the diffusion speeds for different products, and these have found empirical support for diffusion speeds to vary by different prices or investments, user utility or profitability, and the potential market environment of products (Bayus, 1992; Clark et al., 1984; Fisher and Pry, 1971; Griliches, 1957; Mansfield, 1961; 1989; Olshavsky, 1980; Talukdar et al., 2002; Tellis et al., 2003; Van den Bulte, 2000). For example, Mansfield (1961) explained differences in diffusion speeds across industry products using a simple linear model composed of profitability, investment, and variables related to industry environment. Van den Bulte (2000) confirmed that variation in diffusion speeds across consumer durables can be explained by product price and market environment. However, these studies took a static view of diffusing products and did not consider the effect of improving technological performance during the diffusion process on the diffusion speed.

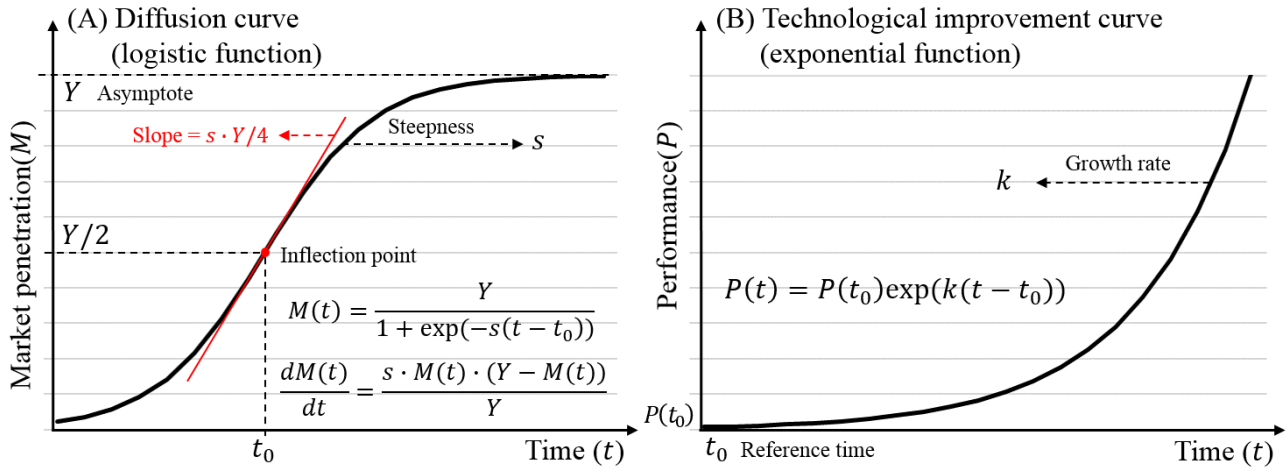

**Figure 1.** Schematic curves of innovation diffusion and technological improvement.
Note1: The logistic function is the most widely used sigmoid function for modeling the S-curve of new product diffusion, but the Gompertz function is also used. The logistic and Gompertz functions have numerical and visual similarities as describing the S-curve of diffusion, but the logistic curve is symmetric and the Gompertz curve is asymmetric (Gompertz curve attains its maximum growth rate at an earlier stage and maintains a more nearly constant growth rate later on, than logistic function); Note2: The steepness parameter determines the steepness of rising logistic curve and the maximum growth rate (speed) of the curve occurs at the inflection point and can be calculated by using the asymptote and steepness parameters.

The diffusion of new products in the potential market follows the S-curve described above, whereas the improvement of a technological domain of a product over time is known to increase exponentially, as shown in the following equation (1) and schematically in Figure 1B for a typical technological domain that might be associated with the product in Figure 1A. The rate of technological improvement varies across technological domains (Farmer and Lafond, 2016; Koh and Magee, 2006; 2008; Magee et al., 2016; Moore, 1965; Nagy et al., 2013; Sahal, 1979).

$$P_i(t) = P_i(t_0)\exp(k_i(t - t_0)) \tag{1}$$

Where $P_i(t)$ and $P_i(t_0)$ represent technological performance at time $t$ and at a reference time $t_0$, respectively, for technological domain $i$, the exponential constant $k_i$ denotes the relative change of performance per year and can be referred to as the technological improvement rate. Since performance is usually measured by output divided by price or other constraints such as volume (this is the inverse of quality adjusted price that is often used as an economic parameter.), the utility or attractiveness of the product to all consumers is increased over time in accordance with equation (1).

Technological improvements are largely time-based, and the diffusion of a product will be affected by the improvement rate of its core technological domain. In this study, we present a regression model to conduct an empirical test for the theoretically expected relationship of diffusion speed with technological improvement rate by extending Mansfield's and other studies' test models as the following equation (2) (Bayus, 1992; Griliches, 1957; Mansfield, 1961; 1989; Olshavsky, 1980; Van den Bulte, 2000). In the test model, we assume that the technological improvement rate as well as the characteristics and potential market environment of the new product identified in previous empirical studies, are important factors in explaining the variation in the diffusion speeds for different products. We use the log-linear regression approach in this study because our dependent variable (diffusion speed) takes only nonnegative values and the distribution is skewed to the right.

$$\ln(S_i) = f\left(N_i, C_i, \frac{d \ln P_i(t)}{dt}\right) = m + \alpha N_i + \beta C_i + \gamma \left(\frac{d \ln P_i(t)}{dt}\right) = m + \alpha N_i + \beta C_i + \gamma k_i \tag{2}$$

Where $S_i$ represents the diffusion speed (the change in penetration per unit time) of product $i$. $N_i$ is a dummy variable indicating a type of market where product $i$ spreads (households or another market). $C_i$ represents the price level of product $i$. Lastly, $k_i$ is the improvement rate of the product's core technological domain.

Based on the above theoretical discussion and the test model, our first hypothesis suggests the relationship of technological improvement with diffusion speeds as follows:

- **Hypothesis H1.** New products that are based on faster-improving technological domains are spread more rapidly in potential markets (i.e., in the test model, $\gamma > 0$; new product diffusion speed $S_i$ is expected to increase with their core technological improvement rates $k_i$).

Some recent studies in technological progress argue that key technical characteristics of technological domains determine the rates of technological improvement, rather than contextual reasons such as investment in research and development (R&D) and organizational aspects (Basnet and Magee, 2016; Benson and Magee, 2015). If we extend this to the theoretical frameworks of the diffusion process described above, we see that the differences in improvement rates of technological domains give different intensity stimuli to demand for related new products, because technological domains have different improvement rates according to their fundamental technical differences.

There are also theoretical studies that discuss the diffusion process through the interaction of supply and demand factors, and these studies suggest another path for how incremental technological improvements might come about and affect the diffusion process (Agarwal and Bayus, 2002; Metcalfe, 1981; Stoneman and Ireland, 1983). Agarwal and Bayus (2002) in particular show that firm entry into a new market is more powerfully correlated than price reductions to sales takeoff of consumer and industrial products. They interpret this result that as a new product is first released into the potential market, it usually takes a primitive form and the demand for it is low; but as firms enter the new market in the diffusion process, non-price competition among firms such as R&D directed towards technological improvements intensifies, which causes firms to launch new products with improved technologies to differentiate themselves, which subsequently increases the demand. This is consistent with our hypothesis H1, in that it emphasizes the significant influence of technological improvement on demand growth at the initial stage of diffusion process. However, this differs from the above-mentioned theoretical framework in that the competition is a major driver of technological improvement rather than technical characteristics of the domain.

Based on the discussion so far, an empirical test to determine which of these two theoretical possibilities can better explain the relationship presented in hypothesis H1 is set up by hypothesis H2. According to the latter theoretical framework, technological performance improves rapidly due to intense competition among firms in the initial stage of the diffusion process, then in the later stage of diffusion process, the technological performance improvement pace is reduced as the competition relaxes, but if the rate of technological improvement is determined by the technical characteristics of the technology domain as suggested in the former theoretical framework, then technological improvement will continue to increase regardless of the competitive dynamics in the diffusion process. Therefore, hypothesis H2 is set as follows:

- **Hypothesis H2.** In the early stage of the diffusion process, the technological performance increases at a faster rate, but the rate of technological improvement decreases in the later stage of the diffusion process.

## 3. Data and methods

*3.1. Innovation diffusion*

*3.1.1. Defining and measuring diffusion speed*

As noted earlier, previous diffusion studies indicate that innovation diffusion follows a S-curve function over time such as the logistic function and define overall diffusion speed using the steepness parameter of this function (parameter $s$ of logistic function shown in Figure 1A) (Griliches, 1957; Mansfield, 1961; 1989; Olshavsky, 1980). This measure of diffusion speed has been widely used in previous studies to identify factors explaining the variation in the diffusion speeds for different products. However, there are a number of problems which can increase the statistical noise greatly in an empirical test of the relationship between technological improvement rates and diffusion speeds. First, this measure of diffusion speed can only assess an average diffusion speed of the entire diffusion process, so such a measure cannot examine changes in diffusion speed over time. Moreover, to use the steepness parameter of the logistic function as a comparable measure of diffusion speed, it is necessary to reliably estimate all aspects of the logistic function. However, if there is insufficient data for the entire diffusion process, or if the diffusion data deviate somewhat from the logistic function, the function cannot be estimated reliably. In addition, the steepness parameter is associated with diffusion speed in reaching the maximum penetration (closely related to the asymptote parameter $Y$ of logistic function shown in Figure 1A), so if the level of maximum penetration among innovations is different, it is difficult to say that the diffusion speed is measured on the same basis (Van den Bulte and Lilien, 1997; Van den Bulte, 2004). That is, to construct a comparable and reliable measure of diffusion speed, start and end points of the diffusion process need to be defined and measured consistently.

Therefore, this study defines diffusion speed as the difference from one penetration level (start point) to a higher level (end point) divided by the penetration time. When we use this kind of measure for diffusion speed, we do not need to estimate speed and maximum penetration parameters of the logistic model and can avoid over-restrictive parametric specification. Moreover, the diffusion speed can be defined according to the penetration interval. Start and end points of the penetration intervals are defined in terms of penetration levels observed directly from data and applied consistently across all products. Specifically, we use the diffusion speed from 0% (market entry) to 10%, the diffusion speed from 10% to 30%, and the diffusion speed from 30% to 50% as three measured diffusion speeds. These diffusion speeds provide insights on both early and later stages of diffusion process, which may relate differently to the rate of technological improvement.

*3.1.2. Diffusion speed data*

In this study, we collect United States penetration or adoption data for 18 innovative products[1] from various sources. The penetration data consists of annual observations of the proportion of consumers who own a particular product among all potential consumers. The data include products launched on the market from the early 1900s to the late 1990s. Fifteen of these are products spreading to households, 1 is a product spreading to farms and 2 are products spreading to hospitals; all of which reached penetration level of more than 50%. The diffusion speed of these products is measured in terms of the change in penetration per year from the market entry to 10% penetration, from 10% to 30%, and from 30% to 50% according to the definition of diffusion speed described above. The diffusion data used in this study are summarized in Table 1.

---

[1] Previous studies finding factors to explain variation in the diffusion speeds for different products used about 10-30 cases because of difficulties in collecting product adoption data. The empirical tests of this study are based on 18 products for which we were also able to find quantitative performance improvement data for the technological domains required to perform core functions of these products. The 18 products included home appliances, consumer electronics, automobiles, and medical imaging equipment. We were also able to include various innovative products related to the diffusion process by which individuals and institutions in a society adopt new technologies or replace older technologies with newer ones, and this includes technologies that have brought major advances and changes to the world. The 18 cases are numerous enough to reliably use linear regression models between two variables (diffusion speed and technological improvement rate) Such models generally require about 10-15 cases to obtain reliable estimates and we conduct non-parametric statistical hypothesis tests for hypothesis H2 for which 18 cases is also sufficient.

**Table 1.** Summary of diffusion data.

| Product | Measure | Year of market entry | Speed from A to B (%/year) | | | Source |
|---|---|---|---|---|---|---|
| | | | 0→10% | 10→30% | 30→50% | |
| 1. Automobile | % of Households | 1898 | 0.59 | 3.33 | 4.00 | Cox and Alm (1997) |
| 2. Washing Machine | % of Households | 1904 | 0.38 | 0.77 | 2.22 | Cox and Alm (1997) |
| 3. Refrigerator | % of Households | 1918 | 0.77 | 3.33 | 4.00 | Cox and Alm (1997) |
| 4. Home Air Conditioning | % of Households | 1929 | 0.36 | 1.67 | 4.00 | Cox and Alm (1997) |
| 5. Dishwasher | % of Households | 1912 | 0.20 | 2.00 | 0.80 | Cox and Alm (1997) |
| 6. Clothes Dryer | % of Households | 1936 | 0.53 | 1.82 | 2.86 | Cox and Alm (1997) |
| 7. Videotape Recorder | % of Households | 1965 | 0.53 | 6.67 | 10.00 | Euromonitor (2017) |
| 8. Personal Computer | % of Households | 1975 | 0.91 | 2.00 | 5.00 | Euromonitor (2017) |
| 9. Laptop | % of Households | 1981 | 0.53 | 3.33 | 5.00 | Euromonitor (2017) |
| 10. Mobile Phone | % of Households | 1983 | 0.91 | 5.00 | 6.67 | Euromonitor (2017) |
| 11. CD Player | % of Households | 1983 | 1.67 | 5.00 | 10.00 | Euromonitor (2017) |
| 12. Internet | % of Households | 1989 | 1.67 | 5.00 | 10.00 | Euromonitor (2017) |
| 13. Digital Camera | % of Households | 1990 | 0.91 | 6.67 | 10.00 | Miranda and Lima (2013) |
| 14. Tablet | % of Households | 1994 | 1.25 | 4.00 | 4.00 | Euromonitor (2017) |
| 15. DVD Player/Recorder | % of Households | 1997 | 3.33 | 10.00 | 10.00 | Euromonitor (2017) |
| 16. Tractor | % of Farms | 1903 | 0.40 | 1.25 | 2.86 | Olmstead and Rhode (2001) |
| 17. Computerized Tomography(CT) scan | % of Hospitals | 1973 | 2.50 | 3.33 | 4.00 | Hillman and Schwartz (1985), Comin and Hobijn (2009), OECD (2017) |
| 18. Magnetic Resonance Imaging(MRI) | % of Hospitals | 1980 | 1.11 | 3.33 | 2.22 | Comin and Hobijn (2009), OECD (2017) |

*3.2. Technological improvement*

*3.2.1. Defining and measuring technological improvement rate*

In this study, Products perform a specific generic function based on a technological domain which is a particular, recognizable body of scientific knowledge (Magee et al. 2016). In this study, we define metrics of the generic functions for core technological domains of products which have the factors affecting adoption decision of products (Magee et al. 2016), in order to assess the rate of technological improvement related to new product diffusion.

We assume that such technological performance metrics follow the exponential function over time as equation (1). Previous studies have empirically confirmed that the exponential relationship between technological performance and time and that the percentage change of performance per year is constant (Benson and Magee, 2015; Farmer and Lafond, 2016; Koh and Magee, 2006; 2008; Magee et al., 2016; Moore, 1965; Nagy et al., 2013; Sahal, 1979), that is consistent with the most widely assumed mechanism of invention-combinatorial progress (Basnet and Magee, 2016; Youn et al., 2015). For example, Magee et al. (2016) showed the strong exponential correlation of performance with time in 28 technological domains and Farmer and Lafond (2016) showed deeper and consistent results for 57 domains. Thus, this study defines the exponential constant $k_i$ (i.e. relative change per year) of equation (1) as the technological improvement rate.

*3.2.2. Technological improvement rate data*

We define performance metrics for core technological domains of products included in our diffusion data and collect the performance data from Magee et al. (2016). Magee et al. (2016) present technical performance trend data in 71 metrics for the 28 technological domains collected from a variety of sources, and the metrics are typically expressed per unit of dollar paid or per unit of other constraints such as volume (the results yield similar yearly improvement rates ($k$) for the different constraints). The performance data and the method used to collect the data is described in detail in Magee et al. (2016). Table 2 reports 11 technological domains matched with 18 diffusing innovative products, including the estimated yearly rate of improvement $k$ and the $R^2$ of the exponential fit; two examples of technological performance over time are plotted in Figure 2. The high value of $R^2$ of the exponential fit in Table 2 indicates that the estimated yearly rate of improvement $k$ well summarizes the observed progress of technological improvement.

**Table 2.** Technological performance data.

| Product | Core technological domain | Metric | Data range (N) | Improvement rate ($k$) | $R^2$ |
|---|---|---|---|---|---|
| 1.Automobile | Piston Engine | Amount of energy produced per unit cost (W/$) | 1896-1971 (22) | 7.12% | 0.79 |
| 2.Washing Machine, 3.Refrigerator, | Electrical Motor | Amount of energy produced per weight (W/kg) | 1881-1993 (13) | 2.93% | 0.83 |
| 4.Home Air Conditioning, 5.Dishwasher, 6.Clothes Dryer | | | | | |
| 7.Videotape Recorder 8.Personal Computer, 9.Laptop, 14.Tablet | Magnetic Information Storage (tape) | Amount of memory per unit cost (Mbits/$) | 1952-2004 (14) | 24.56% | 0.84 |
| | Microprocessor | The number of transistors per die (#/die) | 1972-2006 (12) | 36.33% | 0.97 |
| 10.Mobile Phone | Wireless Telecommunication | Amount of data transfer per second (Kbps) | 1946-2009 (15) | 25.99% | 0.84 |
| 11.CD Player, 15.DVD Player/Recorder | Optical Information Storage | Amount of memory per cc (Mbits/cc) | 1981-2004 (16) | 27.15% | 0.95 |
| 12.Internet | Electrical Telecommunication (Internet Backbone) | Amount of data transfer per second (Kbps) | 1965-2004 (11) | 35.93% | 0.90 |
| 13.Digital Camera | Camera Sensitivity | Sensitivity (mV/micron2) | 1987-2008 (11) | 15.56% | 0.99 |
| 16.Tractor | Tractor Engine | Amount of energy produced per gallon (HP-hr/gallon) | 1920-1964 (17) | 2.77% | 0.84 |
| 17.CT | CT scan | Resolution per scan time (1/resolution·scantime) | 1971-2006 (13) | 36.72% | 0.78 |
| 18. MRI | MRI | Resolution per time per unit cost (1/resolution·scantime·$) | 1980-2006 (6) | 47.52% | 0.88 |

Source: Magee et al. (2016).

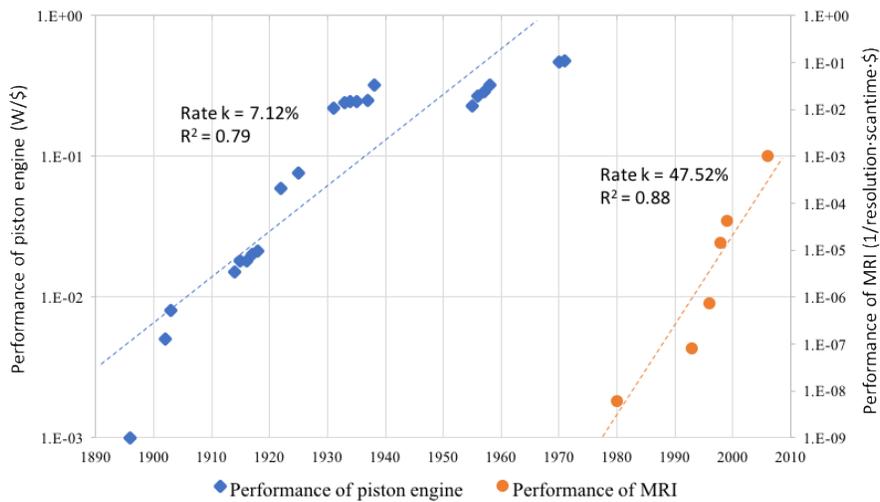

**Figure 2.** Technological performance (logarithmic scale) over time.

## 4. Analysis results

### 4.1. Correlation analysis

First, to examine the relationship between technological improvement and diffusion and to test hypothesis H1, we analyze the correlation between product diffusion speed and technological improvement rate and graph the scatter plots between them. The correlation analysis results are shown in Table 3 and the scatter plots are shown in Figure 3.

**Table 3.** Correlation analysis results.

|  | Mean | S.D. | Min | Max | (1) | (2) | (3) | (4) | (5) |
|---|---|---|---|---|---|---|---|---|---|
| (1) Improvement rate $k$ | 20.78 | 15.68 | 2.77 | 47.52 | 1 | | | | |
| (2) Speed from 0 to 10% | 1.03 | 0.82 | 0.20 | 3.33 | 0.5287 | 1 | | | |
| (3) Speed from 10 to 30% | 3.81 | 2.31 | 0.77 | 10.00 | 0.3823 | 0.6678 | 1 | | |
| (4) Speed from 10 to 50% | 4.34 | 2.59 | 1.14 | 10.00 | 0.3694 | 0.6091 | 0.9662 | 1 | |
| (5) Speed from 30 to 50% | 5.42 | 3.18 | 0.80 | 10.00 | 0.3374 | 0.4920 | 0.8314 | 0.9435 | 1 |

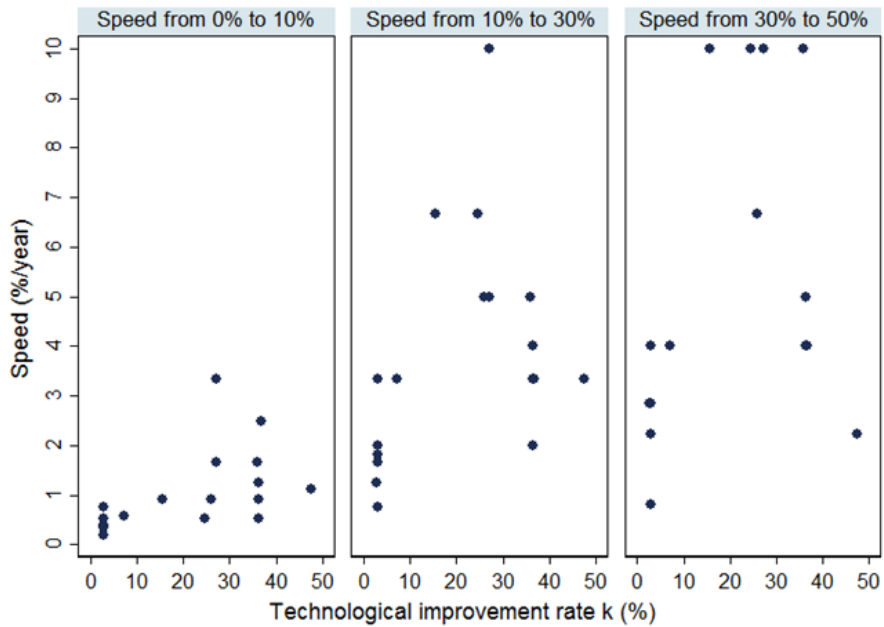

**Figure 3.** Scatter plots between diffusion speed and technological improvement rate.

According to the correlation analysis results and scatters plots, there is a positive correlation between product diffusion speed and technological performance improvement rate. In particular, they show a stronger positive correlation in the early stage (from 0% to 10% penetration) than in the later stage of diffusion process. In other words, the strength of the relationship between product diffusion speed and product improvement rate decreases gradually from the early stage to the later stage of diffusion process. This is expected since the effect of the performance improvement on diffusion is tempered at the later stages by saturation effects. Therefore, both the sign and the fall-off of the correlation with penetration are consistent with our hypothesis H1.

*4.2. Regression analysis*

Next, we empirically test the relationship of diffusion speeds with technological improvement using equation (2), a regression model proposed in this study. In the model, the product diffusion speed from 0% to 10%, from 10% to 30%, and from 30% to 50% are used as dependent variables, respectively. Then, the relative rate of technological improvement, types of market where products spread (household or another market), and level of price are used as explanatory variables. Specifically, products 1 – 15 in Table 1 are set as household adoption products, and products 1 – 10 among these household adoption products are classified as high-priced and the remaining products are classified as low-priced (constant 2015 $1000 is the arbitrary cut-off price between high priced and low priced, and price is based on a rough estimate of the average price in constant 2015 dollars within the product and across time) (Olshavsky, 1980). Table 4 reports the regression analysis results. Each column in Table 4 presents the estimates from a different model. The basic models 1-1, 2-1, and 3-1 have product diffusion speed from 0% to 10%, from 10% to 30%, and from 30% to 50% as dependent variable, respectively, and contain only Improvement Rate k as

an explanatory variable. The models 1-2, 2-2, and 3-2 include additional explanatory variables to control product type and price level in each basic model.

**Table 4.** Regression analysis results.

|  | Model 1-1 (Speed from 0 to 10%) | Model 1-2 (Speed from 0 to 10%) | Model 2-1 (Speed from 10 to 30%) | Model 2-2 (Speed from 10 to 30%) | Model 3-1 (Speed from 30 to 50%) | Model 3-2 (Speed from 30 to 50%) |
|---|---|---|---|---|---|---|
| Improvement rate $k$ | 0.0311*** (0.0086) | 0.0213** (0.0081) | 0.0210** (0.0089) | 0.0176* (0.0088) | 0.0166 (0.0100) | 0.0136 (0.0098) |
| No Household adoption (farms, hospitals) |  | -0.4479 (0.3413) |  | -0.8941** (0.3686) |  | -1.0490** (0.4113) |
| High priced (among household adoption) |  | -0.8246** (0.2800) |  | -0.5812* (0.3024) |  | -0.6092* (0.3374) |
| Constant | -0.8701*** (0.2218) | -0.1345 (0.3119) | 0.7176*** (0.2291) | 1.2616*** (0.3370) | 1.1567*** (0.2575) | 1.7317*** (0.3759) |
| $R^2$ | 0.4487 | 0.6596 | 0.2592 | 0.5000 | 0.1473 | 0.4327 |

Note: the numbers in parentheses are the standard errors of the estimated coefficients.
*significant at the 0.10 level; ** significant at the 0.05 level; *** significant at the 0.01 level.

According to the regression analysis results, Improvement Rate $k$ is in the expected direction for all models and is significant in the models 1 and 2. That is, technological improvement rates are significantly related to diffusion speeds in the early stage of the diffusion process; diffusion speed tends to become faster as technological improvement rates increase. In addition, Both slope and $R^2$ increase for these regressions as we proceed to smaller penetration levels (models 1-1 > 2-1 > 3-1 and 1-2 > 2-2 > 3-2). This result suggests that the relationship between technological improvement rates and diffusion speeds is stronger at the beginning of the diffusion process. Thus, hypothesis H1 is strongly supported; new products that are based on faster-improving technology domains spread more rapidly in potential markets. In addition, new products spreading across farms or hospitals are found to spread relatively slower than products spreading among households. Moreover, high priced household products tend to spread slower in the market.

*4.3. Statistical sign test*

We examine if the rate of technological improvement decreases in the later stage of diffusion process or not, to test our hypothesis H2. First, the diffusion process of products and their technological improvement are graphically plotted over time as shown in Figure 4. These graphs do not show qualitative decrease in the rate of technological improvement in the later stage of diffusion process.

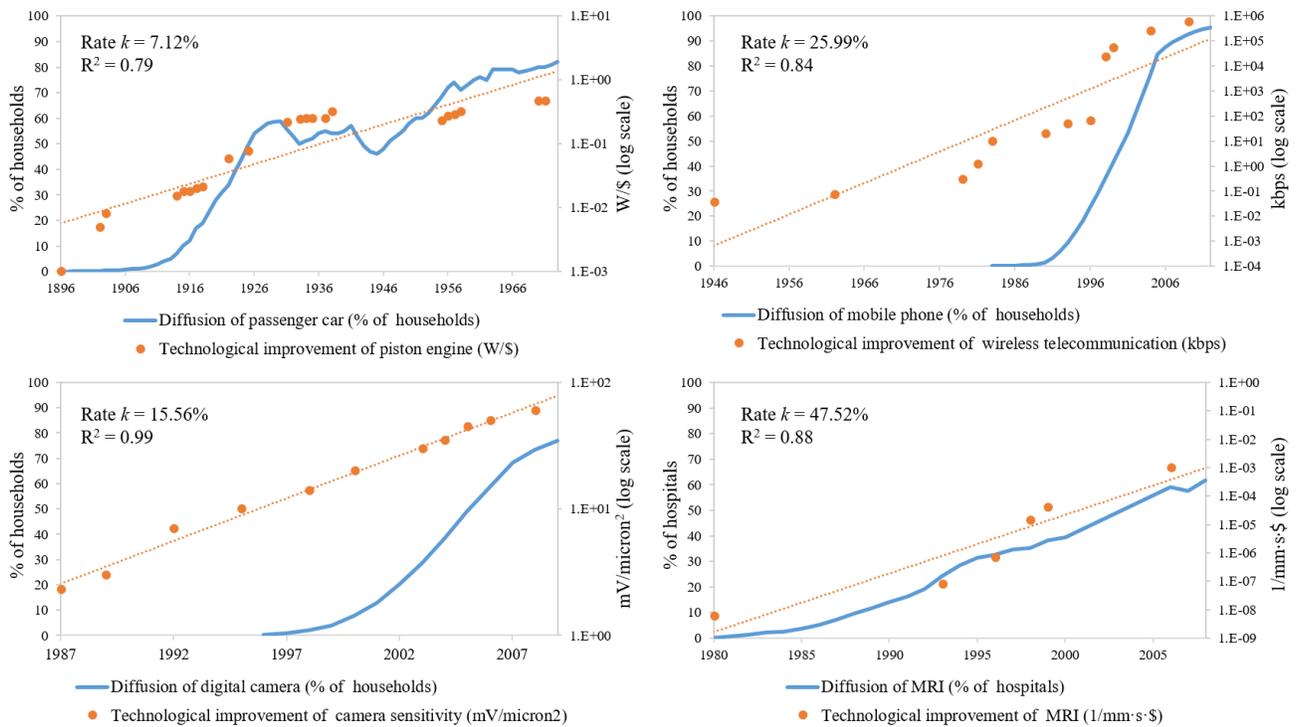

**Figure 4.** Diffusion and technological improvement (logarithmic scale) over time.

Next, the sign test and the Wilcoxon signed rank test (Wilcoxon, 1945) which are statistical methods to test for consistent differences between pairs of observations, are performed to statistically test our hypothesis H2. The null hypothesis of these two tests is that there is no difference between the rates of technological performance of the early and later stages of diffusion process. Then, the alternative hypothesis of the Wilcoxon signed rank test is that there is a difference between them, whereas the alternative hypothesis of the sign test is set up that there is a negative difference between these (i.e., the rate of technological improvement decreases in the later stage of the diffusion process). If the null hypothesis is rejected and the alternative hypothesis is accepted in both tests, we can say that our hypothesis H2 is statistically accepted. Table 5 reports difference in the technological improvement rate between the early and later stages of diffusion process for the statistical tests and the results of the tests.

**Table 5.** Difference in the technological improvement rate between the early and later stages of diffusion.

| Product | Core technological domain (metric) | Early stage (until 10%) | | Later stage (until 50%) | | Difference ($k_2-k_1$) |
|---|---|---|---|---|---|---|
| | | Time range | Improvement rate $k_1$ (%) | Time range | Improvement rate $k_2$ (%) | |
| 1. Automobile | Piston Engine (W/$) | 1896-1915 | 12.96 | 1896-1925 | 12.65 | -0.31 |
| 2. Washing Machine | Electrical Motor (W/kg) | 1881-1929 | 5.50 | 1881-1966 | 3.52 | -1.98 |
| 3. Refrigerator | Electrical Motor (W/kg) | 1881-1929 | 5.50 | 1881-1940 | 5.00 | -0.50 |
| 4. Home Air Conditioning | Electrical Motor (W/kg) | 1881-1964 | 3.90 | 1881-1979 | 3.10 | -0.80 |
| 5. Dishwasher | Electrical Motor (W/kg) | 1881-1966 | 3.52 | 1881-1993 | 2.93 | -0.59 |
| 6. Clothes Dryer | Electrical Motor (W/kg) | 1881-1964 | 3.90 | 1881-1979 | 3.10 | -0.80 |
| 7. Videotape Recorder | Magnetic Information Storage (Mbits/$) | 1952-1984 | 10.53 | 1952-1994 | 14.41 | +3.88 |
| 8. Personal Computer | Microprocessor (#/die) | 1972-1985 | 38.18 | 1972-1999 | 31.79 | -6.39 |
| 9. Laptop | Microprocessor (#/die) | 1972-1999 | 31.79 | 1972-2006 | 36.33 | +4.54 |
| 10. Mobile Telephone | Wireless Telecommunication (Kbps) | 1946-1993 | 15.49 | 1946-2001 | 18.8 | +3.31 |
| 11. CD Player | Optical Information Storage (Mbits/cc) | 1981-1990 | 34.3 | 1981-1995 | 25.73 | -8.57 |
| 12. Internet | Electrical Telecommunication (Kbps) | 1965-1994 | 32.15 | 1965-2002 | 32.8 | +0.65 |
| 13. Digital Camera | Camera Sensitivity (mV/micron2) | 1987-2000 | 16.62 | 1987-2006 | 15.84 | -0.78 |
| 14. Tablet | Microprocessor (#/die) | 1972-2001 | 32.37 | 1972-2006 | 36.33 | +3.96 |
| 15. DVD Player/Recorder | Optical memory (Mbits/cc) | 1981-2000 | 24.25 | 1981-2004 | 27.15 | +2.90 |
| 16. Tractor | Tractor Engine (HP-hr/gallon) | 1920-1929 | 3.87 | 1920-1954 | 3.92 | +0.05 |
| 17. CT | CT (1/mm·s) | 1971-1976 | 180.84 | 1971-1985 | 84.51 | -96.33 |
| 18. MRI | MRI (1/mm·s·$) | 1980-1996 | 26.80 | 1980-2006 | 47.52 | +20.72 |
| **1) Sign test** Null hypothesis: $k_2-k_1 = 0$ vs. Alternative hypothesis: $k_2-k_1 < 0$ p-value = 0.4073 | | | | **2) Wilcoxon signed rank test** Null hypothesis: $k_2-k_1 = 0$ vs. Alternative hypothesis: $k_2-k_1 \neq 0$ p-value = 0.9133 | | |

Note: we also conducted the sign test and the Wilcoxon signed rank test using the data excluding technological domains like electrical motor and microprocessor which match with multiple products to clearly identify the relationship between diffusion process and technological performance progress, but the implication of those results was the same as with the presented results.

As the p-values of the sign test and the Wilcoxon signed rank test turn out to be 0.4073 and 0.9133 respectively and are greater than the 0.05 significance level, we cannot reject the null hypothesis and accept the alternative hypotheses, suggesting that our hypothesis H2 is not supported and there is no observed difference between the rates of technological performance of the early and later stages of diffusion process.

## 5. Discussion and implications

The results based on 18 products and their core technological domains and the associated statistical tests of our hypotheses provide answers to the research questions examined in this study. First, it is expected that the technological improvement rate affects the relative speed of diffusion of the products according to the previous theoretical discussion in the diffusion literature (Chow, 1967; Davies, 1979; Karshenas and Stoneman, 1993). To statistically test this theoretically expected relationship, we set up our hypothesis H1 that new products based on faster-improving technological domains are spread more rapidly in a potential market. Our correlation and regression analysis results confirm that this hypothesis is significantly supported by empirical evidence. Our first finding adds to the limited empirical evidence on the relationship of technological improvement with diffusion showing a broad effect of faster performance improvement leading to faster diffusion.

Additionally, from our correlation and regression analysis results, we find that the intensity of the relationship between technological improvement and diffusion becomes weaker toward the later stage of the diffusion process. It is quite possible that the entry of saturation effects diminish the effect of improved performance. However, it might also be interpreted that the technological improvement gives a greater stimulus to the demand of the related product in the early stage of diffusion process than in the later stage. We can speculate that this occurs because innovators and early adopters accepting a new product in the early stage of the diffusion process are enthusiastic about new technology and high performance, but the majority adopters and laggards accepting the product in the later stage of the diffusion process place more emphasis on low price and stability of the product as Rogers (1995) argues.

Second, according to the existing theoretical argument, the underlying cause of technological improvement and its effect on diffusion can be explained by 1) key technical characteristics of technological domains (Basnet and Magee, 2016; Benson and Magee, 2015) or 2) firm entry and competition among firms in the diffusion process (Agarwal and Bayus, 2002; Metcalfe, 1981; Stoneman and Ireland, 1983). According to the first framework, the performance of the product improves continually at a rate determined by its technical characteristics and it stimulates the demand for the product. On the other hand, according to the second framework, firm entry and competition among firms in the early stage of the diffusion process cause R&D directed towards technological improvement of the product which similarly increases the demand for the product. According to this explanation, the rate of technological improvement of the product decreases at the late stage of the diffusion process as firm entry and competition are reduced. In order to determine which of these two theoretical possibilities can better explain the relationship of technological improvement with diffusion, our hypothesis H2 was set up that the rate of technological improvement slows down in the later stage of the diffusion process as expected from the second framework when new entrants decrease and the competition among firms relaxes. Our statistical sign tests rejected hypothesis H2 and indicate that there is no difference between the rates of technological performance of the early and the later stages of the diffusion process. That is, the results provide empirical evidence that the second framework is not a likely explanation for the observed effect in H1.

Our study provides empirical evidence for the theoretically expected relationship that more rapid diffusion is due to more rapid increases in technical performance, but some might want to consider the opposite direction for causation- more rapid diffusion for any reason leads to more rapid increases in performance improvement. This argument could be based, for example, upon "production learning" leading to faster improvement (Balasubramanian and Lieberman, 2010; Dutton and Thomas, 1984; Lieberman, 1984). However, the results here indicating that performance improvement does not fall off at the later stages of diffusion (or even after diffusion is complete) is a counter-argument to this hypothesis. Moreover, the results in Funk and Magee (2015) and Magee et al. (2016) are even stronger counter-arguments to reversing the direction of causation based upon production since those works show that time rather than production is the key determinant of technological improvement and this is consistent with such rates being determined mostly by fundamental technical factors mediating the general (wide spillover) exponential creation of improvement opportunities (Basnet and Magee, 2016).

Based on our accumulated knowledge to this point, we can summarize that the technological domains of products continue to improve at different rates according to their technical characteristics and the differences in improvement rates of technological domains of products give different intensity stimuli to demand for the products. Thus, new products that are based on faster (slower) improving technological domains spread more rapidly (slowly) in a potential market. However, such relationships weaken when the markets become saturated, even though the technological domains of products continue to improve at specific rates during the diffusion process. In such cases, diffusion of much improved versions of the initial products occur but this is either not counted as diffusion or accounted for by diffusion of a newly named product (desktops-laptops-handheld smart phone, etc.).

## 6. Conclusions

This paper reviews the theoretical basis and empirically examines 18 products and their related technological domains to explore the relationship of technological improvement with innovation diffusion and its underlying cause. Two key findings emerge from our empirical analyses:

- We find that new products that are based on faster (slower) improving technological domains are spread more rapidly (slowly) in a potential market. Moreover, the intensity of the relationship between technological improvement and diffusion becomes weaker toward the later stage of the diffusion process.
- We find that there is no difference between the rates of technological performance of the early and later stages of the diffusion process. This result can be interpreted as: technological domains of products continue to improve at different rates according to their technical characteristics regardless of the diffusion progress, and the differences in improvement rates of technological domains of products give different intensity stimuli to the demand for the products.

## Acknowledgments

The authors gratefully acknowledge the support of the SUTD/MIT International Design Center.